\documentstyle[12pt]{article}

\def\beq{\begin{equation}}
\def\eeq{\end{equation}}
\def\barr{\begin{eqnarray}}
\def\earr{\end{eqnarray}}

\begin{document}

\title{Slater Decomposition of Laughlin States\footnote{UCONN-93-4}}

\author{Gerald V. Dunne \\
Department of Physics\\
University of Connecticut\\
2152 Hillside Road\\
Storrs, CT 06269 USA\\
   \\
dunne@hep.phys.uconn.edu \\}

\date{14 May, 1993}

\maketitle

\begin{abstract}
The second-quantized form of the Laughlin states for the fractional quantum
Hall effect is discussed by decomposing the Laughlin wavefunctions into the
$N$-particle Slater basis. A general formula is given for the expansion
coefficients in terms of the characters of the symmetric group, and the
expansion coefficients are shown to possess numerous interesting symmetries.
For expectation values of the density operator it is possible to identify
individual dominant Slater states of the correct uniform bulk density and
filling fraction in the physically relevant $N\to\infty$ limit.
\end{abstract}

\section{Introduction}
\label{sec-intro}

The variational trial wavefunctions introduced by Laughlin \cite{Lau1} form the
basis for the theoretical understanding of the quantum Hall effect
\cite{Pra1,Sto1}. The Laughlin wavefunctions describe especially stable
strongly correlated states, known as {\it incompressible quantum fluids}, of a
two dimensional electron gas in a strong magnetic field. They also form the
foundation of the hierarchy structure of the {\it fractional} quantum Hall
effect \cite{Hald,Halp,Jain}.

In the extreme low temperature and strong magnetic field limit, the single
particle electron states are restricted to the lowest Landau level. In the
absence of interactions, this Landau level has a high degeneracy determined by
the magnetic flux through the sample \cite{Land}. Much (but not all) of the
physics of the quantum Hall effect may be understood in terms of the
restriction of the dynamics to fixed Landau levels \cite{Gir1,Sto1,Mart}. The
integer quantum Hall effect may be understood in terms of fully filled Landau
levels, but the fractional quantum Hall effect involves fractionally filled
Landau levels. Laughlin's (unnormalized) wavefunctions \cite{Lau1}
\beq
\Psi^{m}_{\rm Laughlin} (z_1,\ldots,z_N) = \prod_{i<j}^{N} (z_i-z_j)^{2m+1}
{}~e^{-{1\over 2}\sum_{i=1}^{N} |z_i|^2} ,\
\label{lw}
\eeq
where $m$ is an integer, correspond to states of fractional filling $1\over
{2m+1}$ in the lowest Landau level. By exploiting the connection with two
dimensional one-component plasmas, it can be shown that these wavefunctions
correspond to incompressible quantum fluids of uniform density ${1\over 2m+1}
\left ({eB\over hc}\right )$ \cite{Lau2}. Furthermore, quasiparticle
excitations about these states have the appropriate fractional statistics
\cite{Arov,Sto1}.

A convenient description of the macroscopic properties of quantum Hall samples
is given by effective field theory techniques and the associated Chern-Simons
formalism \cite{Gir2,Zhan,Sto1}. These effective field theories are in turn
related to important {\it edge} effects in quantum Hall samples \cite{Wen1}.
The second-quantized effective field theories provide a useful formalism for
studying the large $N$ limit. The relationship between the microscopic
many-electron theory and these macroscopic effective field theories has been
discussed in detail for fully filled states by Stone \cite{Sto2}. It is much
more difficult to make precise this connection for the {\it fractional} quantum
Hall effect.

In this paper I discuss the second quantized form of the fractional filling
Laughlin wavefunctions (\ref{lw}) in their microscopic Fock space form. In
practice, this involves expanding the Laughlin wavefunctions in terms of Slater
wavefunctions. Since the Laughlin wavefunction has fixed angular momentum
$J_{\rm Laughlin} = (2m+1){1\over 2}N(N-1)$, it may be expanded as a linear
combination of Slater determinant wavefunctions each with this same angular
momentum. The goal is to seek some structure in this expansion and use this to
examine the $N\to\infty$ limit.

A related question has been addressed very recently by MacDonald and Mitra
\cite{Mitr} who compute the angular momentum distribution function of the
Laughlin wavefunctions - i.e. they compute the relative occupation numbers of
the {\it single particle} angular momentum states in the one particle density
matrix for the Laughlin states. Here, instead, I ask for the relative weights
of the {\it multi-particle} Slater states which comprise the Laughlin state.
This is a much more difficult question, as one sees simply by counting the
states involved. It is often the case that one can compute matrix elements of
second quantized operators in the Slater states with relative ease, so that if
one knows the expansion of the Laughlin state in the Slater basis, one can then
compute the expectation value in the Laughlin state without resorting to the
first quantized technique of large multi-dimensional integrals.

There have been previous investigations of the expansion of the Laughlin
wavefunctions in the Slater basis. Tao \cite{Tao1} has studied the projection
of the Laughlin states on the ground state in the $N\to\infty$ limit. Datta and
Ferrari \cite{Datt} have discussed the large $N$ limit by considering an
associated Langevin equation. As is perfectly clear, any {\it exact} analysis
soon confronts the problem that it is very difficult to store, expand and
manipulate large polyomials and determinants. Here I propose to make use of the
equivalence of this expansion problem with a classic combinatorial problem in
the theory of symmetric functions to give a formula for the coefficients in the
expansion in terms of characters of the symmetric group. The intended advantage
of this approach is that one does not need to refer to the wavefunctions at all
in the computation. There are still significant combinatorial and computational
difficulties, but this point of view brings to light many interesting
symmetries of the

This paper is organized as follows. In Section \ref{sec-secondquant}, I briefly
review the second quantized formalism of $N$-particle quantum mechanics in the
context of the Laughlin wavefunctions. The connection between the many body
Slater wavefunctions and special symmetric polynomials known as "Schur
functions" is discussed in Section \ref{sec-schur}, and this is used in Section
\ref{sec-coeff} to obtain a general formula for the Slater expansion
coefficients in terms of the characters of the symmetric group. Section
\ref{sec-examples} contains the results for some low $N$ examples and in
Section \ref{sec-symmetries} I discuss some symmetries of the expansion
coefficients for all $N$. These symmetries are used in Section
\ref{sec-density} to analyze the density profiles and pair correlation
functions of Laughlin states in the $N\to\infty$ limit. The paper concludes
with some discussion and some suggestions for further investigation.

\section{Second Quantized Form of Laughlin States}
\label{sec-secondquant}
For very large numbers of particles ($N\to \infty$) it is often more
convenient to use the second-quantized formulation of nonrelativistic
quantum mechanics. In second quantization, one defines a field {\it operator}
$\Phi (z)$ by the expansion
\beq
\Phi(z) = \sum _{k=0}^{\infty} a_k \phi_k (z)
\label{fieldop}
\eeq
where the $\phi _k (z)$ form a complete orthonormal set of single particle
wavefunctions and
the $a_k$ are Grassmann annihilation operators satisfying the
anticommutation relations
\beq
\left\{a_k,a_l^{\dag }\right\}=\delta _{kl} \
\label{ccrs}
\eeq
The Fock vacuum $|0>$ is defined as the state annihilated by all the $a_k$
operators
and an N-
particle state $|[\lambda ]>$ is defined by the action of N creation operators
on the vacuum
\barr
|[\lambda ]> & \equiv & |[\lambda_1,\lambda_2, \ldots,\lambda_N]> \nonumber \\
 & \equiv &a_{\lambda_1}^{\dag}~a_{\lambda_2}^{\dag} \ldots
a_{\lambda_N}^{\dag}~|0> \
\label{fock}
\earr
The usual N-particle first-quantized wavefunction corresponding to this state
is
\barr
\Psi ^{[\lambda]} ~(z_1,\ldots,z_N) &\equiv&{1\over \sqrt N!} <0|\Phi(z_1)
\ldots \Phi(z_N)~|[\lambda]> \nonumber\\
&=& {1\over \sqrt N!} \left| \begin{array}{cccc}
\phi_{\lambda_1}(z_1) &\phi_{\lambda_2}(z_1) &\ldots &\phi_{\lambda_N}(z_1)  \\
\phi_{\lambda_1}(z_2) &\phi_{\lambda_2}(z_2) &\ldots &\phi_{\lambda_N}(z_2) \\
\vdots &\vdots & &\vdots \\
\phi_{\lambda_1}(z_N) &\phi_{\lambda_2}(z_N) &\ldots &\phi_{\lambda_N}(z_N)
\end{array}
\right|
\label{slater}
\earr
This is, of course, just the familiar Slater determinant for the N particle
fermionic
state in which the single particle states
$\lambda_1,\lambda_2,\ldots,\lambda_N$ are occupied.
By virtue of (\ref{ccrs}) and (\ref{fock}), these states are orthonormal
\beq
<[\lambda] ~|~ [\mu]>=\delta _{[\lambda]~,~[\mu]} \
\label{ortho}
\eeq
The second quantized density operator is
\beq
\rho (z) = \Phi^{\dag} (z)~\Phi(z) \
\label{densityop}
\eeq
and the (conserved) number operator is
\beq
N=\int d^2 z ~ \rho (z) \
\label{numberop}
\eeq
In the Fock state $|[\lambda]>$, the expectation value of the density operator
is
\beq
< [\lambda]~|~\rho (z) ~|~[\lambda]> = \sum_{i=1}^N |\phi_{\lambda_i} (z) |^2 \
\label{densityexp}
\eeq
The expectation value of the pair correlation operator $\rho(z_1) \rho(z_2)$ is
considered in Section \ref{sec-density}. Notice that, in contrast to the first
quantized approach, there are {\it no integrations}
involved in computing these expectation values.

A general state is a linear combination of the Fock states $|[\lambda]>$
\beq
|state> = \sum_{[\lambda]} c_{~[\lambda]} |[\lambda]>
\label{genstate}
\eeq
and if the coefficients $c_{~[\lambda]}$ are known then it is also
straightforward to
compute expectation values. For example, for the density operator only the
diagonal matrix elements contribute and
\barr
<state|\rho (z)|state> &=&\sum _{[\lambda]} |c_{~[\lambda]}|^2 <[\lambda]|\rho
(z)|[\lambda]>
\nonumber \\
&=&\sum _{[\lambda]} |c_{~[\lambda]}|^2 \left(\sum_{i=1}^N |\phi_{\lambda_i}
(z) |^2 \right)
\label{denstateexp}
\earr
Note, of course, that the coefficients $c_{~[\lambda]}$ uniquely determine the
{\it state}
and have nothing to do with the particular operator. Once the expansion
coefficients are known
for a given state, one can express the expectation value of any operator as a
sum over its Slater state matrix elements. The advantage of the Slater basis is
that these matrix elements are often easy to compute.

Since the first quantized Laughlin wavefunctions are of such fundamental
importance in the
study of the quantum Hall effect, it is natural to ask about their second
quantized form. It is
the purpose of this paper to address this question~:~namely, to determine the
expansion coefficients
$c_{~[\lambda]}$ for the Laughlin state $|~L~>$
\beq
|~L~>=\sum_{[\lambda]} c_{~[\lambda]} |[\lambda]> \
\label{laugh}
\eeq

The corresponding first quantized wavefunction is given by the relation
(\ref{slater}), and so the
$c_{~[\lambda]}$ are determined by the decomposition of the Laughlin
wavefunction (\ref{lw}) in
terms of the Slater basis wavefunctions.

The relevant single particle wavefunctions are those corresponding to the
lowest Landau level (we use
the symmetric gauge for the vector potential $A_i =-{B\over 2}\epsilon_{ij}
x^j$) :
\beq
\phi _k (z) = {1\over \sqrt{\pi k!}}~z^k~e^{-{1\over 2}|z|^2} \
\label{landau}
\eeq
where $k=0,1,2,\ldots$ is an angular momentum label, and where we have used
$\sqrt{2} \ell$ as the unit of length, where $\ell=\sqrt{{\hbar c\over eB}}$ is
the magnetic length. With these single particle wavefunctions, the normalized
Slater wavefunctions (\ref{slater}) are
\barr
\Psi^{[\lambda]}_{\rm Slater} (z_1,z_2,\ldots z_N) = {e^{-{1\over 2}
\sum_{i=1}^{N} |z_i|^2}\over \sqrt{N! \pi^N
\prod_{i=1}^{N} \lambda_i ! }}
\left| \begin{array}{cccc}
z_1 ^{\lambda_1} &z_1 ^{\lambda_2} &\ldots &z_1 ^{\lambda_N}  \\
z_2 ^{\lambda_1} &z_2 ^{\lambda_2} &\ldots &z_2 ^{\lambda_N} \\
\vdots &\vdots & &\vdots \\
z_N ^{\lambda_1} &z_N ^{\lambda_2} &\ldots &z_N ^{\lambda_N}
\end{array}
\right|
\label{lambdaslater}
\earr
There are two convenient simplifications of these wavefunctions. First, it
proves helpful to label the wavefunction not by the string of integers
$[\lambda]=[\lambda_1,\lambda_2,\ldots \lambda_N]$, but by the string
$\{\mu\}=\{\mu_1,\mu_2,\ldots \mu_N\}$ where
\beq
\lambda_i = \mu_i + N- i \
\label{partitions}
\eeq
This simply amounts to labelling relative to the {\it minimum} angular momentum
wavefunction which has $[\lambda]=[N-1,N-2,\ldots,2,1,0]$ and angular momentum
$J={1\over 2 } N(N-1)$. This notation shall be used throughout the paper :
square-bracketed integer strings $[\lambda]$ label the angular momenta of the
occupied single particle states, while the curly-bracketed integer strings
$\{\mu\}$ label states {\it relative} to the state
$[\lambda]=[N-1,N-2,\ldots,2,1,0]$.

The second simplification arises because it is natural to consider the Gaussian
exponential factor $e^{-{1\over 2} |z|^2}$ in the wavefunctions (\ref{landau})
as part of the Hilbert space measure, in which case the lowest Landau level
wavefunctions form a basis for Bargman's Hilbert space of analytic functions
\cite{Gir1,Dun1}. Thus, one may consider the simplified Slater wavefunctions
(denoted by a lower case $\psi$)
\beq
\psi^{\{\mu\}}_{\rm Slater} (z_1,\ldots ,z_N) =
\left| \begin{array}{cccc}
z_1 ^{\mu_1 +N-1} &z_1 ^{\mu_2 +N-2} &\ldots &z_1 ^{\mu_N}  \\
z_2 ^{\mu_1 +N-1} &z_2 ^{\mu_2 +N-2} &\ldots &z_2 ^{\mu_N} \\
\vdots &\vdots & &\vdots \\
z_N ^{\mu_1 +N-1} &z_N ^{\mu_2 +N-2} &\ldots &z_N ^{\mu_N}
\end{array}
\right| \
\label{muslater}
\eeq
The normalization factors in (\ref{lambdaslater}) are not relevant for the
decomposition problem, and will be reintroduced when computing expectation
values - see Section \ref{sec-density}. In this notation, the (un-normalized)
Laughlin wavefunction is
\beq
\psi_{\rm Laughlin} (z_1,\ldots ,z_N) = \left ( \prod_{i<j}^N (z_i-z_j)
\right)^{2 m +1} \
\label{laughlin}
\eeq
where m is an integer. The problem is to expand $\psi_{\rm Laughlin}$ in terms
of the determinants $\psi_{\rm Slater}^{\{\mu\}}$.

Since the Laughlin wavefunction has definite angular momentum
\beq
J_{\rm Laughlin} = (2 m+1){1\over 2} N(N-1) \
\label{angmom}
\eeq
the expansion of $\psi_{\rm Laughlin}$ in terms of the $\psi_{\rm
Slater}^{\{\mu\}}$ will involve only Slater wavefunctions such that
\beq
\sum_{i=1}^N \mu _i = m N(N-1) \
\label{partlength}
\eeq
Thus, one can think of the label $\{\mu\}$ as a {\bf partition} of $m N(N-1)$.
The corresponding label $[\lambda]$, related to $\{\mu\}$ as in
(\ref{partitions}) is instead a partition of $(2 m+1){1\over 2} N(N-1)$.

It is instructive to consider some simple cases. When $N=2$ and $m=1$
(corresponding to the $1\over 3$ filled Laughlin state),
\barr
\psi_{\rm Laughlin} (z_1,z_2)&=&\left( \left | \begin{array}{cc}
z_1 &1 \\
z_2 &1
\end{array} \right | \right ) ^3 \nonumber \\
& & \nonumber \\
&=&\left | \begin{array}{cc}
z_1 ^3 &1 \\
z_2 ^3 &1
\end{array} \right |
-3 \left | \begin{array}{cc}
z_1 ^2 &z_1 \\
z_2 ^2 &z_2
\end{array} \right | \nonumber \\
& & \nonumber \\
&=&\psi_{\rm Slater}^{\{2\}} (z_1,z_2) -3~\psi_{\rm Slater}^{\{1,1\}}(z_1,z_2)
\label{n=2ex}
\earr

When $N=2$ and $m=2$ (corresponding to the $1\over 5$ filled Laughlin state),
\barr
\psi_{\rm Laughlin} (z_1,z_2)&=&\left( \left | \begin{array}{cc}
z_1 &1 \\
z_2 &1
\end{array} \right | \right ) ^5 \nonumber \\
& & \nonumber \\
&=&\left | \begin{array}{cc}
z_1 ^5 &1 \\
z_2 ^5 &1
\end{array} \right |
-5 \left | \begin{array}{cc}
z_1 ^4 &z_1 \\
z_2 ^4 &z_2
\end{array} \right |
+10 \left | \begin{array}{cc}
z_1 ^3 &z_1 ^2\\
z_2 ^3 &z_2 ^2
\end{array} \right |\nonumber \\
& & \nonumber \\
&=&\psi_{\rm Slater}^{\{4\}} (z_1,z_2) -5~\psi_{\rm
Slater}^{\{3,1\}}(z_1,z_2)+10~\psi_{\rm Slater}^{\{2,2\}}(z_1,z_2)
\label{n=2m=2ex}
\earr

When $N=3$ and $m=1$,
\barr
\psi_{\rm Laughlin} (z_1,z_2,z_3)&=&\left( \left | \begin{array}{ccc}
z_1 ^2&z_1 &1 \\
z_2 ^2&z_2 &1 \\
z_3 ^2&z_3 &1
\end{array} \right | \right ) ^3 \nonumber \\
& & \nonumber \\
&=&\left | \begin{array}{ccc}
z_1 ^6 &z_1 ^3 &1 \\
z_2 ^6 &z_2 ^3 &1 \\
z_3 ^6 &z_3 ^3 &1
\end{array} \right |
-3 \left | \begin{array}{ccc}
z_1 ^6 &z_1 ^2 &z_1 \\
z_2 ^6 &z_2 ^2 &z_2 \\
z_3 ^6 &z_3 ^2 &z_3
\end{array} \right |
-3 \left | \begin{array}{ccc}
z_1 ^5 &z_1 ^4 &1 \\
z_2 ^5 &z_2 ^4 &1 \\
z_3 ^5 &z_3 ^4 &1
\end{array} \right |  \nonumber \\
& & \nonumber \\
&&+6 \left | \begin{array}{ccc}
z_1 ^5 &z_1 ^3 &z_1 \\
z_2 ^5 &z_2 ^3 &z_2 \\
z_3 ^5 &z_3 ^3 &z_3
\end{array} \right |
-15 \left | \begin{array}{ccc}
z_1 ^4 &z_1 ^3 &z_1 ^2 \\
z_2 ^4 &z_2 ^3 &z_2 ^2 \\
z_3 ^4 &z_3 ^3 &z_3 ^2
\end{array} \right |  \nonumber \\
& & \nonumber \\
&=&\psi_{\rm Slater}^{\{4,2\}} (z_1,z_2,z_3) -3~\psi_{\rm
Slater}^{\{4,1,1\}}(z_1,z_2,z_3) -3~\psi_{\rm Slater}^{\{3,3\}}(z_1,z_2,z_3)
\nonumber \\
& & \nonumber \\
&&+6\psi_{\rm Slater}^{\{3,2,1\}} (z_1,z_2,z_3) -15~\psi_{\rm
Slater}^{\{2,2,2\}}(z_1,z_2,z_3)
\label{n=3ex}
\earr
For $N=2$ it is trivial to decompose $\psi_{\rm Laughlin}$ into Slater
determinants, but when $N=3$ even for the $m=1$ Laughlin wavefunction it is not
completely trivial to determine the decomposition (\ref{n=3ex}). It is clear
that to proceed to higher numbers of particles and higher m values a more
systematic approach is needed.

In Section \ref{sec-coeff} a closed form expression is presented for the
(integer) coefficients $a_{\{\mu\}}$ appearing in the expansion
\beq
\psi_{\rm Laughlin} = \sum_{\{\mu\}} a_{\{\mu\}} \psi_{\rm Slater}^{\{\mu\}} \
\label{laughdecomp}
\eeq

This expression involves characters of the symmetric group, which will be
briefly reviewed in the next section.

\section{Slater States and Schur Functions}
\label{sec-schur}

It has already been noted that the Slater wavefunctions (\ref{muslater}) which
appear in the expansion (\ref{laughdecomp}) are labelled by partitions
$\{\mu\}$ of $mN(N-1)$. Each of these Slater wavefunctions is a totally
antisymmetric polynomial of homogeneous degree $(2m+1){1\over 2} N(N-1)$. Each
can therefore be written as the {\it product} of the Vandermonde determinant V
and a totally {\it symmetric} polynomial of homogeneous degree $mN(N-1)$. Here
V is given by
\barr
V&=&\left| \begin{array}{cccc}
z_1 ^{N-1} &z_1 ^{N-2} &\ldots &1  \\
z_2 ^{N-1} &z_2 ^{N-2} &\ldots &1 \\
\vdots &\vdots & &\vdots \\
z_N ^{N-1} &z_N ^{N-2} &\ldots &1
\end{array}
\right| \nonumber \\
&=& \prod_{i<j}^N (z_i-z_j)
\label{van}
\earr
The corresponding symmetric polynomial is called a "Schur function" or
"S-function" \cite{Litt,MacD}.
\beq
{\cal S} ^{\{\mu\}} (z_1,\ldots ,z_N) \equiv {{\left| \begin{array}{cccc}
z_1 ^{\mu_1 +N-1} &z_1 ^{\mu_2 +N-2} &\ldots &z_1 ^{\mu_N}  \\
z_2 ^{\mu_1 +N-1} &z_2 ^{\mu_2 +N-2} &\ldots &z_2 ^{\mu_N} \\
\vdots &\vdots & &\vdots \\
z_N ^{\mu_1 +N-1} &z_N ^{\mu_2 +N-2} &\ldots &z_N ^{\mu_N}
\end{array}
\right|}\over {\left| \begin{array}{cccc}
z_1 ^{N-1} &z_1 ^{N-2} &\ldots &1  \\
z_2 ^{N-1} &z_2 ^{N-2} &\ldots &1 \\
\vdots &\vdots & &\vdots \\
z_N ^{N-1} &z_N ^{N-2} &\ldots &1
\end{array}
\right|}} \
\label{schur}
\eeq
These symmetric polynomials play a key role in the theory of the symmetric
group. This relationship between the Slater states and the Schur functions has
previously been exploited for fully filled QHE states by Stone \cite{Sto2},
while the suggestion of characterizing the quantum hall effect wavefunctions in
terms of symmetric polynomials was originally made by Laughlin \cite{Lau1}.

The unfamiliar reader should pause to note that it is by no means obvious at
first sight that the ratio of determinants in (\ref{schur}) produces a
polynomial!

For the purposes of this paper, the most important property of the Schur
functions ${\cal S}^{\{\mu\}}$ is that they provide a linear basis for the
space of symmetric polynomials of homogeneous degree $\sum_{i=1}^N \mu_i$.
The Laughlin wavefunction (\ref{laughlin}) is just $V^{2m+1}$ and so, dividing
the expansion (\ref{laughdecomp}) through by the Vandermonde determinant V, one
finds
\barr
{\psi_{\rm Laughlin}\over V}&\equiv&V^{2m} \nonumber \\
&=&\sum_{\{\mu\}} a~_{\{\mu\}} {\cal S}^{\{\mu\}}
\label{schurdecomp}
\earr
Thus, to determine the coefficients $a~_{\{\mu\}}$ of the expansion of
$\psi_{\rm Laughlin} \equiv V^{2m+1}$ in terms of the totally antisymmetric
Slater wavefunctions $\psi_{\rm Slater}^{\{\mu\}}$, equivalently one may
consider the expansion of $V^{2m}$ in terms of the totally symmetric Schur
functions ${\cal S}^{\{\mu\}}$. From (\ref{laughdecomp}) and
(\ref{schurdecomp}), the expansion coefficients are in 1-1 correspondence. This
reduction of degree of the polynomial under consideration from $V^{2m+1}$ to
$V^{2m}$ represents a significant computational simplification.

Another convenient basis for the space of totally symmetric polyomials of
homogeneous degree $mN(N-1)$ is the "power sum basis". The power sums $s_\mu$
are defined for $\mu=1 \ldots \infty$ as
\beq
s_\mu (z_1,\ldots ,z_N) \equiv \sum_{i=1}^N ~z_i ^\mu \
\label{powersum}
\eeq
Any totally symmetric polynomial may be expressed as a sum of products of the
$s_\mu$'s, and so a basis for the totally symmetric polynomials of homogeneous
degree $mN(N-1)$ is provided by the products
\beq
s_{\{\mu\}} \equiv s_{\mu_1} s_{\mu_2} \ldots s_{\mu_N}\
\label{powerbasis}
\eeq
where $\sum_{i=1}^N \mu_i=mN(N-1)$.

The linear transformation between the Schur function basis and the power sum
basis is given by Frobenius' reciprocity formula \cite{Litt,MacD}:
\beq
s_{\{\mu\}} = \sum_{\{\lambda\}} \chi_{~\{\mu\}}^{\{\lambda\}} {\cal
S}_{\{\lambda\}} \
\label{frobenius}
\eeq
where the $\chi_{~\{\mu\}}^{\{\lambda\}}$ are characters of the representation
$\{\lambda\}$ of the symmetric group of $\sum_{i=1}^N \mu_i$ symbols. For the
application in this paper, $\sum_{i=1}^N \mu_i=mN(N-1)$, and so the relevant
symmetric group is $S_{mN(N-1)}$.

It is important to note that the characters $\chi_{~\{\mu\}}^{\{\lambda\}}$ are
{\bf integers}. Furthermore, these characters satisfy the following
completeness relations \cite{Litt,MacD} which permit the inversion of
Frobenius' formula (\ref{frobenius}). If the partition $\{\mu\}$ of $mN(N-1)$
is labelled with the Frobenius notation $(l_1,l_2,l_3,...)$ where
\beq
l_1 +2l_2 +3l_3 + \ldots = mN(N-1) \
\label{lpart}
\eeq
then the characters $\chi_{~\{\mu\}}^{\{\lambda\}} \equiv
\chi_{~(l)}^{\{\lambda\}}$ satisfy
\barr
{1\over \left(mN(N-1)\right) !}\sum_{(l)} g_{(l)}
\chi_{~(l)}^{\{\lambda\}}\chi_{~(l)}^{\{\lambda '\}} &=&
\delta_{\{\lambda\},\{\lambda '\}} \nonumber \\
\sum_{\{\lambda\}} \chi_{~(l)}^{\{\lambda\}}\chi_{~(l')}^{\{\lambda
\}}&=&{\left( mN(N-1)\right)!\over g_{(l)}}~\delta_{(l),(l')}
\label{complete}
\earr
where
\beq
g_{(l)} \equiv {{\left(mN(N-1)\right) !}\over {1^{l_1}2^{l_2}\ldots l_1 !l_2 !
\ldots}} \
\eeq

\section{Formula for the Expansion Coefficients $a_{\{\mu\}}$}
\label{sec-coeff}

{}From Frobenius' formula (\ref{frobenius}), it is clear that if one can expand
$V^{2m}$ in the power sum basis, it is then straightforward to convert this
exansion to the Schur function basis. This is a useful observation because it
is relatively easy to express $V^{2m}$ in the power sum basis since the
"discriminant" $V^2$ may be expanded as
\barr
V^2&=&\left| \begin{array}{cccc}
1 &1 &\ldots &1  \\
z_1 &z_2 &\ldots &z_N \\
\vdots &\vdots & &\vdots \\
z_1 ^{N-1} &z_2 ^{N-1} &\ldots &z_N ^{N-1}
\end{array}
\right|
\left| \begin{array}{cccc}
1 &z_1 &\ldots &z_1 ^{N-1}  \\
1 &z_2 &\ldots &z_2 ^{N-1} \\
\vdots &\vdots & &\vdots \\
1 &z_N &\ldots &z_N ^{N-1}
\end{array}
\right| \nonumber \\
& & \nonumber \\
&=&\left| \begin{array}{ccccc}
N &s_1 &s_2 &\ldots &s_{N-1}  \\
s_1 &s_2 &s_3 &\ldots &s_N \\
s_2 &s_3 &s_4 &\ldots &s_{N+1} \\
\vdots &\vdots & &\vdots \\
s_{N-1} &s_N &s_{N+1} &\ldots &s_{2(N-1)}
\end{array}
\right| \nonumber \\
& & \nonumber \\
&=& \sum _{perms.~p} (-1)^p~s_{p(1)-1} s_{p(2)} s_{p(3)+1} \ldots
s_{p(N)+N-2}
\label{vsquare}
\earr
where in the last sum, the permutations p are permutations on $N$ letters, and
$s_0 \equiv N$.

Thus, equation (\ref{vsquare}) provides a simple decomposition of $V^2$ in the
power sum basis
\beq
V^2 = \sum_{perms.~p} (-1)^p s_{\{\mu_p \}} \
\label{vsquaredecomp}
\eeq
where for each permutation p, $\{\mu_p \}$ is the (unordered) partition
\beq
\{\mu_p \} \equiv \{p(1)-1,p(2),p(3)+1,\ldots ,p(N)+N-2 \} \
\label{muppartitions}
\eeq
of $N(N-1)$. $V^{2m}$ may similarly be expanded in terms of the power sum
basis, and then in the Schur function basis by using Frobenius' formula
(\ref{frobenius}). For definiteness and simplicity, we shall henceforth
concentrate on the $m=1$ case, which corresponds to the Laughlin wavefunction
for fractional filling $1\over 3$.

In the definition (\ref{powersum}) of the power sums, it is assumed that $\mu
\geq 1$, so that the partitions in (\ref{powerbasis}) are partitions into {\it
nonzero parts}. Before applying Frobenius' formula to (\ref{vsquaredecomp}) it
is necessary to separate the Laplace expansion (\ref{vsquare}) into
permutations for which $s_0$ is a factor (i.e those $p$ for which $p(1)=1$) and
those for which $s_0$ is not a factor. The former correspond to partitions of
$N(N-1)$ into exactly $N-1$ nonzero parts, while the latter correspond to
partitions of $N(N-1)$ into exactly $N$ nonzero parts.
(This is a drastic reduction compared to {\it all possible} partitions of
$N(N-1)$). Then, by Frobenius' formula, one finds the following expression for
the expansion coefficients $a~_{\{\mu\}}$:
\beq
a_{\{\mu\}} = N \sum_{p~s.t.~p(1)=1} (-1)^p \chi_{~\{\lambda_p \}}^{\{\mu\}}
+\sum_{p~s.t.~p(1)\neq 1} (-1)^p \chi_{~\{\lambda_p \}}^{\{\mu\}} \
\label{coeff}
\eeq
It is important to stress a computational issue at this point. This formula
allows one to compute the coefficient of any Schur function in the expansion of
$V^2$ without ever having to make any reference to the {\it polynomial} $V^2$
itself. This is an enormous computational simplification, as the algebraic
manipulation of large polynomials rapidly becomes prohibitively difficult. The
characters $\chi_{~\{\lambda\}}^{\{\mu\}}$ may be computed efficiently using
(for example) the combinatorial package "combinat" on Maple V \cite{Mapl}.

In the next section the results for $N=2,3,4,5,6$ are presented. It is worth
noting that although the expression (\ref{coeff}) for $a_{\{\mu\}}$ is a
(large) sum of integers, the final answers for the $a_{\{\mu\}}$ are
particularly simple integers, with many interesting symmetry properties not
evident from the expansion
(\ref{coeff}).

\section{Examples}
\label{sec-examples}

Before proceeding to the statement and discussion of some computer-generated
results, it is instructive to consider in detail some low $N$ examples to see
how the formula (\ref{coeff}) works. For $N=2$,
\barr
V^2 &=&\left | \begin{array}{cc}
2 &s_1 \\
s_1 &s_2
\end{array} \right | \nonumber \\
&=&2 s_2 - s_1 ^2 \nonumber \\
&\equiv& 2 s_{\{2\}} - s_{\{1,1\}}
\label{n=2power}
\earr
The character table for the symmetric group $S_2$ is shown in Table
\ref{S2char}.

Frobenius' formula (\ref{frobenius}) says
\barr
s_{\{2\}} = {\cal S}_{\{2\}} -{\cal S}_{\{1,1\}} \nonumber \\
s_{\{1,1\}} = {\cal S}_{\{1,1\}} +{\cal S}_{\{2\}} \nonumber \\
\earr
which one may easily check since the explicit $N=2$ power sums and Schur
functions are
\barr
s_{\{2\}} &= &z_1 ^2 + z_2 ^2 \nonumber \\
s_{\{1,1\}} &=&(z_1 +z_2)^2 \nonumber \\
{\cal S}_{\{2\}}&\equiv&{\left | \begin{array}{cc}
z_1 ^3 &1 \\
z_2 ^3 &1
\end{array} \right |} \left / {\left | \begin{array}{cc}
z_1 &1 \\
z_2 &1
\end{array} \right |} \right. \nonumber \\
&=&z_1 ^2 + z_1 z_2 +z_2 ^2 \nonumber\\
{\cal S}_{\{1,1\}}&\equiv&{\left | \begin{array}{cc}
z_1 ^2 &z_1 \\
z_2 ^2 &z_2
\end{array} \right |} \left / {\left | \begin{array}{cc}
z_1 &1 \\
z_2 &1
\end{array} \right |} \right. \nonumber \\
&=&z_1 z_2 \nonumber\\
\earr

Inserting these expansions into (\ref{n=2power}) yields
\beq
V^2 ={\cal S}_{\{2\}} -3{\cal S}_{\{1,1\}} \
\label{n=2schur}
\eeq
which should be compared with the Slater expansion of $V^3$ in (\ref{n=2ex}).

For $N=3$,
\barr
V^2&=&\left| \begin{array}{ccc}
3 &s_1 &s_2 \\
s_1 & s_2 &s_3 \\
s_2 & s_3 &s_4
\end{array} \right | \nonumber \\
&=& -s_2 ^3 + 2s_1 s_2 s_3 -3 s_3 ^2 - s_1 ^2 s_4 + 3 s_2 s_4 \nonumber\\
&\equiv& - s_{\{2,2,2\}} +2 s_{\{3,2,1\}} -3 s_{\{3,3\}} -s_{\{4,1,1\}} +3
s_{\{4,2\}}
\label{n=3power}
\earr
It is clear that this decomposition of $V^2$ into the power sum basis is
labelled by partitions of $N(N-1)=6$ into $2$ or $3$ parts. The character table
for the symmetric group $S_6$ is $10{\rm x}10$, but in fact only a $5{\rm x}5$
subblock contributes, corresponding to the $5$ partitions appearing in the
expansion (\ref{n=3power}).

Then, from equation (\ref{coeff}),
\beq
a_{\{\mu\}} =3\left(\chi_{~\{4,2\}}^{\{\mu\}}-\chi_{~\{3,3\}}^{\{\mu\}}\right)
+2 \chi_{~\{3,2,1\}}^{\{\mu\}} -\chi_{~\{2,2,2\}}^{\{\mu\}} -
\chi_{~\{4,1,1\}}^{\{\mu\}} \
\label{n=3coeff}
\eeq
Then, using the character table for $S_6$ in Table \ref{S6char} one finds the
expansion
\beq
V^2 = {\cal S}_{\{4,2\}}-3{\cal S}_{\{4,1,1\}}-3{\cal S}_{\{3,3\}}+6{\cal
S}_{\{3,2,1\}}-15{\cal S}_{\{2,2,2\}} \
\label{n=3schur}
\eeq
which once again should be compared with the explicit expansion (\ref{n=3ex})
of $V^3$ into Slater determinants.

For $N=4$ one needs the characters of $S_{12}$ \cite{Jack}. Then Equation
(\ref{coeff}) leads to an expansion of $V^2$ in terms of $16$ different Schur
functions:
\barr
V^2 &=& {\cal S}_{\{6,4,2\}}-3{\cal S}_{\{5,5,2\}}-3{\cal S}_{\{6,3,3\}}+6{\cal
S}_{\{5,4,3\}}\nonumber \\
& &-15{\cal S}_{\{4,4,3\}} -3{\cal S}_{\{6,4,1,1\}}+9{\cal
S}_{\{5,5,1,1\}}+6{\cal S}_{\{6,3,2,1\}}\nonumber \\
& &-12{\cal S}_{\{5,4,2,1\}}-9{\cal S}_{\{5,3,3,1\}}+27{\cal
S}_{\{4,4,3,1\}}-15{\cal S}_{\{6,2,2,2\}}\nonumber \\
& &+27{\cal S}_{\{5,3,2,2\}}-6{\cal S}_{\{4,4,2,2\}}-45{\cal
S}_{\{4,3,3,2\}}+105{\cal S}_{\{3,3,3,3\}}
\label{n=4schur}
\earr

For $N=5$ there are $59$ Schur functions in $V^2$ and for $N=6$ there are
$247$. The results for the $N=5$ expansion are presented in Table
\ref{n=5schur}. The $N=6$ expansion is tabulated in the Appendix. These
expansions have been checked by explicit expansion of the polynomial $V^3$ in
terms of Slater determinants, using Mathematica \cite{Math}, although that
technique deals directly with the polynomials themselves and so is slower.
Notice that the coefficients are 'simple' integers, and there are clear
symmetries and recursive patterns - these will be discussed in more detail in
Section \ref{sec-symmetries}.

\section{Symmetry Properties of Expansion Coefficients}
\label{sec-symmetries}

The most striking symmetry property of the expansion coefficients is the
following {\it exact } symmetry. For a given partition $\{\mu\}$, define the
"reversed" partition $\{\tilde{\mu}\}$ by
\beq
\{\tilde{\mu}\}=\{ 2(N-1)-\mu_N,2(N-1)-\mu_{N-1},\ldots ,2(N-1)-\mu_1\}\
\label{reverse}
\eeq
Then the expansion coefficients of $\{\mu\}$ and $\{\tilde{\mu}\}$ are equal:
\beq
a_{\{\mu\}} = a_{\{\tilde{\mu}\}} \
\label{revcoeff}
\eeq
This result has direct physical significance because the integers $\mu_i$
correspond to the angular momenta $\lambda_i = \mu_i +N -i$ of single particle
states and these states are strongly localized at radius $\sqrt{\lambda_i}$.
In terms of the single particle state labels $[\lambda]$ in (\ref{partitions}),
the reversal operation is
\beq
[\tilde{\lambda}] =
[3(N-1)-\lambda_N,3(N-1)-\lambda_{N-1},\ldots,3(N-1)-\lambda_1] \
\label{lamreverse}
\eeq
 Thus, the relation (\ref{revcoeff}) relates the expansion coefficients of
states peaked at one point in the droplet with those for states peaked at other
points. In particular, this may be used to relate states localized near the
edge of the droplet with those localized near the center.

While this symmetry is clear from the examples presented in the previous
section, it is not at all clear from the expansion coefficient formula
(\ref{coeff}). There is, however, an easy proof\footnote{Thanks to D.~Jackson
for suggesting this approach} of (\ref{revcoeff}). $V^2(z_1,\ldots ,z_N)$ is a
homogeneous polynomial of degree $N(N-1)$ and it satisfies
\barr
V^2(z_1,\ldots ,z_N)&=&\left ( \prod_{i=1}^N z_i^{2(N-1)} \right ) V^2({1\over
z_1},\ldots ,{1\over z_N}) \nonumber \\
&=&\left ( \prod_{i=1}^N z_i^{2(N-1)} \right ) \sum_{\{\mu\}} a_{\{\mu\}} {\cal
S}_{\{\mu\}} \left({1\over z_1},\ldots,{1\over z_N} \right )
\label{inverse}
\earr
But, as can be seen from the definition (\ref{schur}),
\beq
\left ( \prod_{i=1}^N z_i^{2(N-1)} \right ) {\cal S}_{\{\mu\}} \left({1\over
z_1},\ldots,{1\over z_N} \right ) = {\cal S}_{\{\tilde{\mu}\}} \left(
z_1,\ldots,z_N \right )\
\label{schurinverse}
\eeq
Hence,
\barr
V^2(z_1,\ldots ,z_N)&=&\sum_{\{\mu\}} a_{\{\mu\}} {\cal S}_{\{\tilde{\mu}\}}
\left( z_1,\ldots,z_N \right ) \nonumber \\
&=&\sum_{\{\mu\}} a_{\{\tilde{\mu}\}} {\cal S}_{\{\mu\}} \left( z_1,\ldots,z_N
\right )
\label{vsquarereverse}
\earr

In addition to this exact symmetry of the Slater expansion coefficients, it is
possible to deduce simple combinatorial formulas (for {\it any} N) for the
coefficients of certain special Slater states. These are highly nontrivial
results, and somewhat fortuitously correspond to physically important Slater
states in the $N\to\infty$ limit (see Section (\ref{sec-density})).

The simplest such example is the state
\beq
\{\mu\} = \{2(N-1),2(N-2),\ldots,4,2\} \
\label{spread}
\eeq
for which
\beq
a_{\{\mu\}} =1. \
\label{spreadcoeff}
\eeq
This corresponds to the Slater state
\beq
|[\lambda]> = |[3(N-1),3(N-2),\ldots,6,3,0]> \
\label{spreadstate}
\eeq
in which every third single particle angular momentum state is filled. It is
also the most uniformly distributed (in the bulk) of the Slater states (see
Section \ref{sec-density}). The other extreme is the Slater state in which the
angular momentum levels are most closely bunched
\beq
\{\mu\} =\{N-1,N-1,\ldots,N-1\} \
\label{bunched}
\eeq
for which
\beq
a_{\{\mu\}}=\left ( -1 \right ) ^{[N/2]} (2N -1 )!! \
\label{bunchedcoeff}
\eeq
where $[N/2]$ means the integer part of $N/2$. Furthermore, {\bf all} the
expansion coefficients $a_{\{\mu\}}$ lie between these extremes (in magnitude)
\beq
1\leq |a_{\{\mu\}} | \leq (2N-1)!! ~~~~~~{\rm ,~for~all}~\{\mu\}.\
\label{rangecoeff}
\eeq
Each of these states (\ref{spread}) and (\ref{bunched}) is invariant under the
"reversal" operation defined above (\ref{reverse}). However, another important
state (which is not reversal invariant) is that for which one electron is in
the $0$ angular momentum state and the remaining $N-1$ electrons are bunched
together
\beq
\{\mu\} = \{\overbrace{N,N,\ldots,N}^{\rm N-1}\}\
\label{<bunched}
\eeq
for which
\beq
a_{\{\mu\}} = \left (-1\right) ^{[N/2]} (2N-3)!!\
\label{<bunchedcoeff}
\eeq
The reversed partition is
\beq
\{\tilde{\mu}\} = \{2(N-1),N-2,\ldots,N-2\}, \
\label{<bunchedrev}
\eeq
for which, by (\ref{revcoeff}), the coefficient is also given by
(\ref{<bunchedcoeff}).

Other closed formulas exist for the reversal-invariant states that begin with
the "maximally bunched" state (\ref{bunched}) and successively move the extreme
inner and outer electrons in and out (respectively) by one step :
\barr
a_{\{2N,2N-2,2N-3,\ldots,N+1,N,N-2\}}& = &\left (-1\right )^{[N/2]+1}
(N-1)(2N-3)!! \nonumber \\
a_{\{2N+1,2N-2,2N-3,\ldots,N+1,N,N-3\}}& = &\left (-1\right )^{[N/2]+1}
N(N-1)(2N-5)!! \nonumber \\
 &\vdots & \nonumber \\
a_{\{3(N-1),2N-2,2N-3,\ldots,N+1,N,0\}}& = &\left (-1\right )^{[N/2]+1}
(2N-5)!!
\label{closed}
\earr
These all correspond to states highly localized in the region $\sqrt{N} \leq
|z| \leq \sqrt{2(N-1)}$.

Similarly, one may begin with the maximally distributed state (\ref{spread})
and make local shifts of electrons between angular momentum levels. The
simplest such shifts just involve one pair of electrons, in which one electron
is raised by one angular momentum step and another is lowered (remember that
the {\it total} angular momentum must not change). If these two electrons were
initially separated by $3$ units of angular momentum, then after such a shift
they will be separated by just $1$ unit of angular momentum. Consulting the
results for the coefficients $a_{\{\mu\}}$ we note the remarkable fact that
such an operation always changes the expansion coefficient $a_{\{\mu\}}$ by a
factor of $-3$. If the two electrons were initially separated by $6$ units
(with another electron midway in between) then the change in the coefficient
$a_{\{\mu\}}$ is $+6$. In fact, in an outer sub-block of M uniformly spaced
electrons (spaced by 3 units of angular momentum, as in the state
(\ref{spread})), if the outermost an
\beq
(-1)^{M-1} 3~.~2^{M-2} \
\eeq
For example, the state with
\beq
\{\mu\} = \{2N-3,2N-4,2N-6,\ldots,4,2,1\} \
\label{<spreadmu}
\eeq
has angular momentum labels
\beq
[\lambda] = [ 3N-4,3N-6,3N-9,\ldots,6,3,1] \
\label{<spreadlambda}
\eeq
and differs from the maximally distributed state by the bringing together (by
one unit each of angular momentum) of the outermost and innermost electrons,
and its expansion coefficient is
\beq
a_{\{\mu\}} = (-1)^{N-1} 3~.~2^{N-2} \
\label{domcoeff}
\eeq

There are also clear {\it recursive} properties of the expansion coefficients.
For example, the $N$ particle states in which one particle is in the
"innermost" $0$ level are in one-to-one correspondence with {\bf all} the
$(N-1)$ particle states. Further, by the reversal symmetry property
(\ref{revcoeff}), this also applies to the $N$ particle states for which one
particle is in the "outermost" $3(N-1)$ level. This property is a simple
consequence of the fact that one may expand the $N$ particle Vandermonde
determinant in powers of any given $z_i$, say $z_1$,as
\barr
V_N(z_1,z_2,\ldots,z_N) &=& \prod_{1\leq i <j}^N (z_i-z_j) \nonumber \\
&=&V_{N-1}(z_2,\dots,z_N) \prod _{j=2}^N (z_1-z_j) \\
&=&V_{N-1}(z_2,\dots,z_N) \nonumber  \\
&& (z_1 ^{N-1} -e_1 {z_1}^{N-2}+e_2 {z_1}^{N-3}- \dots +(-1)^{N-1} e_{N-1} )
\nonumber
\label{vanexp}
\earr
where $e_\mu =e_{\mu} (z_2,\ldots,z_N)$ is the $\mu ^{\rm th}$ elementary
symmetric polynomial in the $N-1$ variables $z_2,\ldots,z_N$. ($e_\mu$ is
defined to be the sum of all products of $\mu $ of the variables; e.g. $e_1=z_2
+\ldots+z_N$, and $e_{N-1} = z_2 z_3 \ldots z_N$ ). From the expansion
(\ref{vanexp}) it is clear that the only way to obtain the term $z_1 ^{3(N-1)}$
(this corresponds to putting particle $1$ in the outermost level) in $V_N ^3$
is by accompanying it by $\left (V_{N-1} (z_2,\ldots,z_N)\right)^3$, and so one
is left with the decomposition problem for $N-1$ particles.

Similar reasoning leads to further, deeper, recursive patterns. Define $\#(N)$
to be the number of Schur functions appearing in the expansion of $V_N ^2$
(equivalently, the number of Slater determinants appearing in the expansion of
$V_N ^3$). Then, with a consistent ordering of the coefficients (see tables)
one sees that the first $\#(N-1)$ coefficients for $N$ particles coincide with
{\bf all} the coefficients for $N-1$ particles. Further, the {\bf next}
$\#(N-2)$ coefficients of the $N$ particle problem are given by $-3$ times the
$\#(N-2)$ coefficients of the $N-2$ particle problem; the next $\#(N-3)$
coefficients of the $N$ particle problem are given by $6$ times the $\#(N-3)$
coefficients of the $N-3$ particle problem; the next $\#(N-4)$ coefficients of
the $N$ particle problem are given by $-12$ times the $\#(N-4)$ coefficients of
the $N-4$ particle problem,etc$\dots$. For example, for the $N=5$ case, the
first $16+5+2=23$ coefficients are determined by this recursive pattern from
the $N=2,3,4$ resul

Of course, ultimately one would like a simple combinatorial formula (of the
form of (\ref{closed}) for example) for {\it all} the coefficients in the
expansion for {\it any} $N$. No such formula is currently known, although the
results presented here suggest that such a simple formula may exist. Indeed,
the reversal symmetry and recursive properties of the coefficients, together
with their simple nature suggest that such a formula would be a product of
ratios of factorials. However, any such formula would necessarily involve in
general all the parameters $\{\mu\}$ which specify the partition in question -
the simple closed formulas given above are for especially symmetric classes of
partitions.

In the absence of such a general formula for the coefficients $a_{\{\mu\}}$ for
arbitrary $N$, one may still discuss the question of whether the Slater
decomposition (\ref{laughdecomp}) is in some sense "dominated" by certain
special states. This is not yet a well posed question, because the notion of
which states dominate depends on which expectation value is being considered,
and so this is no longer a property of the Laughlin state itself. However, it
is still possible to identify important representative Slater states from the
decomposition (\ref{laughdecomp}) and estimate their relative weights exactly
as $N\to \infty$. Remember, of course, that in order to address such questions,
it is necessary to reintroduce the correct normalization factors (which were
previously dropped for convenience) so that the expansion of the Laughlin state
is in terms of {\it normalized} Slater states, rather than just in terms of the
bare Slater determinants themselves. These issues will be discussed further in
the next Se

\section{Density Profiles}
\label{sec-density}

In this Section, the expansion coefficients for $N=2,3,4,5,6$ are used to plot
the expectation value of the particle density and the pair correlation function
in the $1\over 3$ filled Laughlin state. Allowing for the change in the
labelling convention from $[\lambda] \to \{\mu\}$ as given by
(\ref{partitions}), the expectation value of the density operator is
\beq
<L|\rho (z)|L>={1\over {\cal N}} \sum_{\{\mu\}} |c_{\{\mu\}} |^2 \sum_{i=1}^N
|\phi_{\mu_i +N-i} (z) |^2 \
\label{laughden}
\eeq
where the normalized single particle states are given by
(\ref{landau}),${1\over {\cal N}}$ is an overall normalization factor, and the
coefficients $c_{\{\mu\}}$ are related to the coefficients $a_{\{\mu\}}$ of
(\ref{laughdecomp}) by the reintroduction of the Slater normalization factors
in (\ref{lambdaslater})
\beq
c_{\{\mu\}} = a_{\{\mu\}} \sqrt{ \prod_{i=1}^N (\mu_i +N-i)!} \
\label{ccoeff}
\eeq
The common factors of $\sqrt{N! \pi ^N}$ have been absorbed into the overall
normalization which is fixed by demanding that the integral of the expectation
value of the density operator is just the total particle number
\beq
\int d^2 z <L|\rho (z)|L> = N \
\label{number}
\eeq
This is achieved by taking
\beq
{\cal N} = {1\over \sqrt{N! \pi ^N}} \sum _{\{\mu\}} |a_{\{\mu\}} |^2 \left (
\prod _{i=1}^N (\mu _i +N-i)! \right ) \
\label{norm}
\eeq
Since the size of the $1\over 3$ filled droplet increases with $N$, in order to
compare the density profile for different $N$ it is appropriate to rescale the
length by a factor $\sqrt{3N}$ (for the general Laughlin state of filling
fraction $1\over {2m+1}$, this rescaling factor would be $\sqrt{(2m+1)N}$)
\beq
r\equiv {1\over \sqrt{3 N}} |z| \
\label{rescale}
\eeq
Then one can plot
\beq
\hat{\rho} (r) \equiv \pi <L|\rho (\sqrt{3N} r) |L> \
\label{scaling}
\eeq
This rescaling is chosen so that the {\bf edge} of the droplet is at $r=1$ {\bf
for all N}. The normalization is chosen such that for a {\it perfectly} uniform
$1\over 3$ filled state $\hat{\rho}(r)$ would be a step function
\beq
\hat{\rho}_{\rm uniform}(r) = \left\{ \begin{array}{lll}
                         {1\over 3}& &0\leq r < 1\\
                          0 & &r> 1
                        \end{array}
\right. \
\label{step}
\eeq
In Figure \ref{densityplot} is plotted the exact expectation value of the
density in the ${1\over 3}$ filled Laughlin state for $N=2,3,4,5,6$ particles.
These plots show a clear tendency towards the form (\ref{step}). For these
small values of $N$ there is also evidence of the characteristic "boundary
hump" studied by Datta and Ferrari \cite{Datt} (see also Figure
\ref{humpplot}).

The density profiles (\ref{scaling}) are {\it weighted} sums of the density
profiles of the individual Slater states which make up the Laughlin state.
\beq
\hat{\rho} (r) = \left( {\pi \over \sum_{\{\mu\}} |c_{\{\mu\}}|^2}\right )
\sum_{\{\mu\}} |c_{\{\mu\}}|^2 \hat{\rho}_{\{\mu\}}(r) \
\label{expansion}
\eeq
where the density profile of the individual Slater state $\{\mu\}$ is
\beq
\hat{\rho}_{\{\mu\}}(r) \equiv e^{-3N r^2} \sum_{i=1}^{N} {(3N r^2)^{\mu_i
+N-i} \over (\mu_i +N-i)! } \
\label{slatexp}
\eeq
The {\it relative} weighting factors are just given by the squares of the
expansion coefficients $c_{\{\mu\}}$, with the overall normalization fixed by
(\ref{number}). Note that each density function $\hat{\rho}_{\{\mu\}}(r)$ is
bounded $0\le \hat{\rho}_{\{\mu\}}(r)\le 1$ for all $r$. This means that,
given information about the expansion coefficients, one can ask which (if any)
Slater states dominate this weighted sum in the $N\to\infty$ limit.

Consider, for example, the Slater density for the maximally distributed state,
$[\lambda]=|[3(N-1),3(N-2),\ldots,3,0]>$, for which the $\{\mu\}$ partition is
$\{\mu\}=\{2(N-1),2(N-2),\ldots,2,0\}$ and for which the expansion coefficient
is $a_{\{\mu\}}=1$. Then the weighted expansion coefficient in the density sum
is
\beq
|c_{\{2(N-1),2(N-2),\ldots,2,0\}}|^2 = \prod_{k=0}^{N-1} (3 k)! \
\label{maxcoeff}
\eeq
The density profile of this individual Slater state is given by
\beq
\hat{\rho}_{\{2(N-1),\ldots,2,0\}}(r) = e^{-3N r^2} \sum_{k=0}^{N-1} {(3N
r^2)^{3k} \over (3k)!}\
\label{maxden}
\eeq
This is plotted in Figure \ref{uniplot} for $N=10,100,100$, and one sees that
in the bulk of the droplet, this Slater state (by itself!) represents a uniform
density of fractional filling $1\over 3$ with remarkable accuracy even for such
low values of $N$. (The analogous result is true for all fractional fillings
$1\over{2m+1}$). The failure at $r=0$ and $r=1$ is analogous to the breakdown
of the WKB approximation near classical turning points (see \cite{Cap1,Cap2}
for a discussion of the importance of the large N and semiclassical limits at
the edge of quantum Hall samples). In applications to the quantum Hall effect
one is actually more interested in {\it annular} samples \cite{Lau2,Hal2}
(rather than disc samples) and so $r=0$ is replaced by an edge at some small
nonzero radius.

This is, however, just one term of many contributing to the Laughlin density
expectation value. One should compare it with the other terms in the expansion.
The states with the largest $a_{\{\mu\}}$ expansion coefficients are the
maximally bunched states (\ref{bunched}) with $\{\mu\}=\{N-1,N-1,\dots,N-1\}$,
for which $|a_{\{\mu\}}|=(2N-1)!!$. These states are highly localized near
central values of the radius. These states have Slater occupation numbers given
by (\ref{bunched}) and (\ref{partitions})
\beq
|[\lambda]> = |[2N-2,2N-3,\ldots,N,N-1]> \
\label{bunchedlambda}
\eeq
and, relative to (\ref{maxcoeff}) their weighting in the expansion
(\ref{expansion}) is
\beq
{|c_{\{\mu\}}|^2 \over \prod_{k=0}^{N-1} (3 k)!} = {\left( (2N-1)!!\right) ^2
\left( (2N-2)! (2N-3)! \ldots N! (N-1)! \right) \over \prod_{k=0}^{N-1} (3
k)!}\
\label{ratio}
\eeq
For small $N$ this increases as the numerator dominates. For $N=12$ the ratio
is $\sim 2. 10^6$, but after $N=12$ the denominator begins to dominate and the
ratio decreases rapidly. For $N=20$ the ratio is $\sim 0.2$, and by the time
$N=50$ it has dropped dramatically to $\sim 10^{-134}$. The relative magnitudes
of the $c_{\{\mu\}}$ involves a {\it competition} between the $a_{\{\mu\}}$
factors and the factorial normalization factors $\sqrt{\prod_{i=1}^{N} (\mu_i
+N -i)!}$. From the results in the previous section, the $a_{\{\mu\}}$ factors
tend to increase as the electrons become more closely bunched together. But the
factorial factors tend to increase with the angular momentum levels being as
high as possible. One might therefore expect that the states of the form
(\ref{<bunched}) which have $N-1$ electrons bunched together at high angular
momentum and just one electron in the $k=0$ level to be a dominant state in the
density expansion. From (\ref{<bunched}) and (\ref{<bunchedcoeff}) its relative
wieght i
\beq
{|c_{\{\mu\}}|^2 \over \prod_{k=0}^{N-1} (3 k)!} = {\left( (2N-3)!!\right) ^2
\left( (2N-1)! (2N-2)!  \ldots (N+2)! (N+1)! \right) \over \prod_{k=0}^{N-1} (3
k)!}\
\label{rratio}
\eeq
This ratio also increases with small $N$ , but peaks at $N=15$ at a value of
$\sim 2.10^9$, and then decreases rapidly. For $N=20$ it is still $\sim 10^6$,
but for $N=50$ it is $\sim 10^{-109} $. For low $N$ these are the {\it most
dominant} states in the density expansion, and from Figures \ref{densityplot}
and \ref{humpplot}, in which the density profiles are plotted for
$N=2,3,4,5,6$, one sees their contribution to the characteristic "boundary
hump" discussed previously in \cite{Datt}.

In fact, the factorial factors in (\ref{ccoeff}) eventually dominate in the
$N\to\infty$ limit, since there is an increasing number of {\it factorials} of
numbers which are themselves increasing. This grows much more quickly than the
$a_{\{\mu\}}$ factors, which are bounded in magnitude by $(2N-1)!!$. This tends
to favor the evenly distributed states. Indeed, the dominant states are those
which are "close to" the maximally distributed state (\ref{spread}). By "close
to" is meant states which are related to (\ref{spread}) by a shift of just two
electrons. Shifts of more than two electrons tend to be suppressed by the
factorial factors in the $N\to\infty$ limit. Of these states, the ones with the
largest weighting coefficients $c_{\{\mu\}}$ are those states (\ref{<spreadmu})
in which the innermost electron is raised from the $0$ level to the $k=0$ level
and the outermost electron is lowered from the $k=3(N-1)$ level to the $k=3N-4$
level. For these states,
\barr
{|c_{\{\mu\}}|^2 \over \prod_{k=0}^{N-1} (3 k)!}& = &9~.~2^{2(N-2)} {(3N-4)!
(3N-6)! \ldots 3! 1! \over (3N-3)! (3N-6)! \ldots 3! 0!} \nonumber \\
&=&3 {2^{2(N-2)} \over {N-1}}
\label{domrat}
\earr
This ratio tends to infinity as $N\to\infty$. The density profile of this
dominant class of states is given, for all $N$, by
\beq
\hat{\rho}_{\rm dom}(r) = e^{-3N r^2} \left ( 3N r^2 + \sum_{k=1}^{N-2} {(3N
r^2)^{3k} \over (3k)! } + {(3N r^2)^{3N-4} \over (3N-4)!} \right ) \
\label{domexp}
\eeq
These density profiles are plotted in Figure \ref{domplot} for $N=10,100,1000$.
These plots show that in the bulk of the droplet the uniform density of the
$1\over  3$ filled Laughlin state is well approximated by the density profile
of this individual Slater state. Also, apart from the behavior near $r=0$ these
plots are indistinguishable from those of the maximally distributed state
(\ref{spread}) in Figure \ref{uniplot}.

The results for the expansion coefficients found for $N=2,3,4,5,6$ may also be
used to plot the pair correlation function $g(z_1,z_2)$. Conventionally
\cite{Morf,Lau2,Janc}, one considers the radial pair correlation
\barr
 g(|z|)&\equiv &g(|z|,0) \nonumber \\
&=&{{<L|\rho(|z|) \rho(0)|L>}\over{<L|\rho(|z|)|L><L|\rho(0)|L>}}
\label{gdef}
\earr
near the center of the disc-like sample, to minimize edge effects. Since the
density operator (\ref{densityop}) at the center has the simple form $\rho(0) =
a^{\dag}_0 a_0$, it is easy to see that $g(|z|)$ is given in terms of the
expansion coefficients $c_{\{\mu\}}$ and the single particle states
(\ref{landau}) by
\beq
g(|z|)={{\sum_{\{\mu\}~s.t.~\mu_N \ne 0} |c_{\{\mu\}}|^2 \sum_{i=1}^{N-2}
|\phi_{\mu_i+N-i}(|z|)|^2}\over {\left( \sum_{\{\mu\}} |c_{\{\mu\}}|^2
|\phi_{\mu_i+N-i}(|z|)|^2 \right) \left( \sum_{\{\mu\}~s.t.~\mu_N \ne 0}
|c_{\{\mu\}}|^2 \right)}}
\label{gcorr}
\eeq

In order to compare with previous computations using the first quantized
approach \cite{Morf,Janc,Mac1}, this has been plotted in Figure \ref{corrplot}
for the $N=6$ case in terms of the variable $x={1\over \sqrt{2}}|z|$ (i.e. in
units of the magnetic length, rather than the scaling (\ref{rescale}) used
previously for the density plots). The exact result from Equation (\ref{gcorr})
and the coefficients (\ref{ccoeff}) computed here for the $N=6$ $1\over 3$
filled Laughlin state agrees (as it must) with the exact result of MacDonald
and Murray \cite{Mac1}, which as they point out is very close to the
$N\to\infty$ Monte Carlo result \cite{Cail}. In Figure \ref{corrplot}, I have
also plotted the pair correlation function computed from (\ref{gcorr}) in the
$N\to\infty$ limit by just retaining a {\it single} dominant Slater state in
each of the sums appearing in the expression (\ref{gcorr}). This leads to
\beq
g_{\rm approx}(x)={{\left({x^2 /2}\right)^4 + \sum_{j=2}^{N-2}{\left({x^2
/2}\right)^{3j} \over (3j)!} +{\left({x^2 /2}\right)^{3N-4} \over (3N-4)!}}
\over  {{x^2 /2} + \sum_{j=1}^{N-2}{\left({x^2 /2}\right)^{3j} \over (3j)!}
+{\left({x^2 /2}\right)^{3N-4} \over (3N-4)!}}} \
\label{gapprox}
\eeq
This function (\ref{gapprox}) at $N=6$ is also virtually indistinguishable from
its $N\to\infty$ limit. From the Figure \ref{corrplot} one sees that this
truncated pair correlation function has the correct tendency, but slightly
overestimates $g(x)$ at distances less than $\sim 3$ magnetic lengths. This may
be understood since the use of expression (\ref{gdef}) for the pair correlation
function assumes uniformity near the origin $x=0$, which is not true for the
individual $N\to\infty$ 'dominant' Slater states (see Figures
\ref{uniplot},\ref{domplot}).

\section{Conclusions}
\label{concl}
To conclude, the re-interpretation of the expansion of the Laughlin
wavefunctions in Slater wavefunctions as the expansion of $V^{2m}$ in terms of
Schur functions leads to an expression for the coefficients of this expansion
in terms of the characters of the symmetric group $S_{mN(N-1)}$. This may be
used to peform the Slater decomposition of the Laughlin states without
explicitly expanding the polynomial $V^{2m+1}$. Some low dimensional examples
are presented here (as is seen from the Appendix, even the {\it tabulation} of
results for higher $N$ becomes difficult) and from these it has been possible
to glean some important symmetry properties of the expansion coefficients. In
many cases this means a simple combinatorial formula for the expansion
coefficient. While a simple product formula has not been found for all states
for all $N$, it is still possible to show that for the expectation value of the
density operator there is a single dominant Slater state which has the correct
uniform bulk density and fil

However, the question of the 'dominance' of the Slater expansion of the
Laughlin state is specific to the operator in question. Thus, the dominant
state for the expectation value of the density operator is in no way
necessarily dominant when considering the expectation values of other
operators, such as the pair correlation operator or the energy operator.
Indeed, the state (\ref{<spreadmu}) resembles the superlattice state,
considered originally by Tao and Thouless \cite{Tao2}, which was unable to
explain all features of the fractional quantum Hall effect. The question of
dominance for the density is simplified by the fact that the individual density
profiles $\hat{\rho}$ are bounded between $0$ and $1$. It would be important to
understand this issue for other operators, especially the energy operator.

It would also be interesting to explore the relationship with the approach of
MacDonald and Mitra \cite{Mitr} which concentrates on the one particle density
matrix. They compute the angular distribution function $<n_m>$ which is related
to the expansion coefficients $c_{[\lambda]}$ discussed here by $<n_m> =
\sum_{[\lambda]_m} (c_{[\lambda]})^2$, where the sum $\sum_{[\lambda]_m}$ is
the sum over all $[\lambda]$ such that the angular momentum index $m \in
[\lambda]$.

\bigskip
\bigskip
\bigskip
\bigskip
\bigskip
\bigskip
\noindent{\bf Acknowledgements}

Thanks to D.~Jackson for invaluable advice on combinatorial matters. Thanks
also to A.~Cappelli, C.~Trugenberger and G.~Zemba for helpful discussions. This
work was supported in part by the D.O.E. under Grant DE-FG02-92ER40716.00, and
in part by the University of Connecticut Research Foundation. I am grateful to
J.~Jain for bringing reference \cite{Tao2} to my attention.

\bigskip
\bigskip
\bigskip
\bigskip
\bigskip
\bigskip
\noindent{\bf Note Added}
\bigskip

After completion of this work I became aware of a package called "SF" written
for Maple V by J.~Stembridge \cite{Stem}, which includes procedures for
manipulating and analyzing symmetric functions. One of its procedures converts
a symmetric function expressed in the power sum basis into an expansion in the
Schur function basis. For low $N$, using the analysis of Section
\ref{sec-coeff}, this gives a very efficient way to compute the Slater
decomposition of the Laughlin states. It would be interesting to test the
efficiency of this package for larger $N$.

\newpage
\clearpage

\begin{table}[p]
\centering
\begin{tabular}{|c||r|r|} \hline
$\chi_{~\{\mu\}}^{\{\lambda\}}$ & \{ 1,1 \} & \{ 2 \} \\
\hline\hline
\{ 1,1 \} &1 & -1 \\ \hline
\{ 2 \} &1 &1 \\
\hline
\end{tabular}
\caption{character table for symmetric group $S_2$}
\label{S2char}
\end{table}

\newpage
\clearpage

\begin{table}[p]
\centering
\begin{tabular}{|c||r|r|r|r|r|} \hline
$\chi_{~\{\mu\}}^{\{\lambda\}}$ & \{ 2,2,2 \} & \{ 3,2,1 \} &\{3,3\} &\{4,1,1\}
&\{4,2\} \\
\hline\hline
\{2,2,2 \} &3 &-1 &2 &1 &-1 \\ \hline
\{3,2,1 \}  &1 &1  &-2&0 &0  \\ \hline
\{3,3\}  &-3 &1 &2 &-1 &-1 \\  \hline
\{4,1,1\}  &-2 &-1 &1 &0 &0 \\  \hline
\{4,2\}  &3 &0 &0 &-1 &1 \\  \hline
\end{tabular}
\caption{partial character table for symmetric group $S_6$}
\label{S6char}
\end{table}

\newpage
\clearpage

\begin{table}[p]
\centering
\begin{tabular}{|c|r||c|r||c|r|} \hline
\{$\mu$\}&$a_{\{\mu\}}$  &\{$\mu$\}&$a_{\{\mu\}}$  &\{$\mu$\}&$a_{\{\mu\}}$
\\ \hline
\{   8, 6, 4, 2, 0\}&  1&\{      6, 6, 6, 1, 1\}&  45&\{      8, 4, 4, 2, 2\}&
-6\\
\{    7, 7, 4, 2, 0\}&  -3&\{      8, 6, 3, 2, 1\}&  6&\{      7, 5, 4, 2, 2\}&
 -27\\
\{    8, 5, 5, 2, 0\}&  -3&\{      7, 7, 3, 2, 1\}&  -18&\{      6, 6, 4, 2,
2\}&  111\\
\{    7, 6, 5, 2, 0\}&  6&\{      8, 5, 4, 2, 1\}&  -12&\{      6, 5, 5, 2,
2\}&  -18\\
\{    6, 6, 6, 2, 0\}&  -15&\{      7, 6, 4, 2, 1\}&  24&\{      8, 4, 3, 3,
2\}&  -45\\
\{    8, 6, 3, 3, 0\}&  -3&\{      7, 5, 5, 2, 1\}&  18&\{      7, 5, 3, 3,
2\}&  81\\
\{    7, 7, 3, 3, 0\}&  9&\{      6, 6, 5, 2, 1\}&  -54&\{      6, 6, 3, 3,
2\}&  -18\\
\{    8, 5, 4, 3, 0\}&  6&\{      8, 5, 3, 3, 1\}&  -9&\{      7, 4, 4, 3, 2\}&
 72\\
\{    7, 6, 4, 3, 0\}&  -12&\{      7, 6, 3, 3, 1\}&  18&\{      6, 5, 4, 3,
2\}&  -144\\
\{    7, 5, 5, 3, 0\}&  -9&\{      8, 4, 4, 3, 1\}&  27&\{      5, 5, 5, 3,
2\}&  45\\
\{    6, 6, 5, 3, 0\}&  27&\{      7, 5, 4, 3, 1\}&  -36&\{      6, 4, 4, 4,
2\}&  -90\\
\{    8, 4, 4, 4, 0\}&  -15&\{      6, 6, 4, 3, 1\}&  -27&\{      5, 5, 4, 4,
2\}&  270\\
\{    7, 5, 4, 4, 0\}&  27&\{      6, 5, 5, 3, 1\}&  81&\{      8, 3, 3, 3,
3\}&  105\\
\{    6, 6, 4, 4, 0\}&  -6&\{      7, 4, 4, 4, 1\}&  -36&\{      7, 4, 3, 3,
3\}&  -180\\
\{    6, 5, 5, 4, 0\}&  -45&\{      6, 5, 4, 4, 1\}&  72&\{      6, 5, 3, 3,
3\}&  45\\
\{    5, 5, 5, 5, 0\}&  105&\{      5, 5, 5, 4, 1\}&  -180&\{      6, 4, 4, 3,
3\}&  270\\
\{    8, 6, 4, 1, 1\}&  -3&\{      8, 6, 2, 2, 2\}&  -15&\{      5, 5, 4, 3,
3\}&  -75\\
\{    7, 7, 4, 1, 1\}&  9&\{      7, 7, 2, 2, 2\}&  45&\{      5, 4, 4, 4, 3\}&
 -420\\
\{    8, 5, 5, 1, 1\}&  9&\{      8, 5, 3, 2, 2\}&  27&\{      4, 4, 4, 4, 4\}&
 945\\
\{    7, 6, 5, 1, 1\}&  -18&\{      7, 6, 3, 2, 2\}&  -54& & \\ \hline
\end{tabular}
\caption{Schur function decomposition for N=5}
\label{n=5schur}
\end{table}

\newpage
\clearpage

\noindent {\bf Appendix : N=6 Schur function decomposition}
\vspace{1in}

\begin{table}[h]
\centering
\begin{tabular}{|c|r||c|r||c|r|} \hline
\{$\mu$\}&$a_{\{\mu\}}$  &\{$\mu$\}&$a_{\{\mu\}}$  &\{$\mu$\}&$a_{\{\mu\}}$
\\ \hline
\{10, 8, 6, 4, 2, 0\}&1&\{9, 8, 5, 5, 3, 0\}&18&\{7, 7, 6, 5, 5, 0\}&-75\\
\{9, 9, 6, 4, 2, 0\}&-3&\{10, 6, 6, 5, 3, 0\}&27&\{7, 6, 6, 6, 5, 0\}&-420\\
\{10, 7, 7, 4, 2, 0\}&-3&\{9, 7, 6, 5, 3, 0\}&-36&\{6, 6, 6, 6, 6, 0\}&945\\
\{9, 8, 7, 4, 2, 0\}&6&\{8, 8, 6, 5, 3, 0\}&-27&\{10, 8, 6, 4, 1, 1\}&-3\\
\{8, 8, 8, 4, 2, 0\}&-15&\{8, 7, 7, 5, 3, 0\}&81&\{9, 9, 6, 4, 1, 1\}&9\\
\{10, 8, 5, 5, 2, 0\}&-3&\{9, 6, 6, 6, 3, 0\}&-36&\{10, 7, 7, 4, 1, 1\}&9\\
\{9, 9, 5, 5, 2, 0\}&9&\{8, 7, 6, 6, 3, 0\}&72&\{9, 8, 7, 4, 1, 1\}&-18\\
\{10, 7, 6, 5, 2, 0\}&6&\{7, 7, 7, 6, 3, 0\}&-180&\{8, 8, 8, 4, 1, 1\}&45\\
\{9, 8, 6, 5, 2, 0\}&-12&\{10, 8, 4, 4, 4, 0\}&-15&\{10, 8, 5, 5, 1, 1\}&9\\
\{9, 7, 7, 5, 2, 0\}&-9&\{9, 9, 4, 4, 4, 0\}&45&\{9, 9, 5, 5, 1, 1\}&-27\\
\{8, 8, 7, 5, 2, 0\}&27&\{10, 7, 5, 4, 4, 0\}&27&\{10, 7, 6, 5, 1, 1\}&-18\\
\{10, 6, 6, 6, 2, 0\}&-15&\{9, 8, 5, 4, 4, 0\}&-54&\{9, 8, 6, 5, 1, 1\}&36\\
\{9, 7, 6, 6, 2, 0\}&27&\{10, 6, 6, 4, 4, 0\}&-6&\{9, 7, 7, 5, 1, 1\}&27\\
\{8, 8, 6, 6, 2, 0\}&-6&\{9, 7, 6, 4, 4, 0\}&-27&\{8, 8, 7, 5, 1, 1\}&-81\\
\{8, 7, 7, 6, 2, 0\}&-45&\{8, 8, 6, 4, 4, 0\}&111&\{10, 6, 6, 6, 1, 1\}&45\\
\{7, 7, 7, 7, 2, 0\}&105&\{8, 7, 7, 4, 4, 0\}&-18&\{9, 7, 6, 6, 1, 1\}&-81\\
\{10, 8, 6, 3, 3, 0\}&-3&\{10, 6, 5, 5, 4, 0\}&-45&\{8, 8, 6, 6, 1, 1\}&18\\
\{9, 9, 6, 3, 3, 0\}&9&\{9, 7, 5, 5, 4, 0\}&81&\{8, 7, 7, 6, 1, 1\}&135\\
\{10, 7, 7, 3, 3, 0\}&9&\{8, 8, 5, 5, 4, 0\}&-18&\{7, 7, 7, 7, 1, 1\}&-315\\
\{9, 8, 7, 3, 3, 0\}&-18&\{9, 6, 6, 5, 4, 0\}&72&\{10, 8, 6, 3, 2, 1\}&6\\
\{8, 8, 8, 3, 3, 0\}&45&\{8, 7, 6, 5, 4, 0\}&-144&\{9, 9, 6, 3, 2, 1\}&-18\\
\{10, 8, 5, 4, 3, 0\}&6&\{7, 7, 7, 5, 4, 0\}&45&\{10, 7, 7, 3, 2, 1\}&-18\\
\{9, 9, 5, 4, 3, 0\}&-18&\{8, 6, 6, 6, 4, 0\}&-90&\{9, 8, 7, 3, 2, 1\}&36\\
\{10, 7, 6, 4, 3, 0\}&-12&\{7, 7, 6, 6, 4, 0\}&270&\{8, 8, 8, 3, 2, 1\}&-90\\
\{9, 8, 6, 4, 3, 0\}&24&\{10, 5, 5, 5, 5, 0\}&105&\{10, 8, 5, 4, 2, 1\}&-12\\
\{9, 7, 7, 4, 3, 0\}&18&\{9, 6, 5, 5, 5, 0\}&-180&\{9, 9, 5, 4, 2, 1\}&36\\
\{8, 8, 7, 4, 3, 0\}&-54&\{8, 7, 5, 5, 5, 0\}&45&\{10, 7, 6, 4, 2, 1\}&24\\
\{10, 7, 5, 5, 3, 0\}&-9&\{8, 6, 6, 5, 5, 0\}&270&\{9, 8, 6, 4, 2, 1\}&-48\\
\hline
\end{tabular}
\end{table}

\newpage
\clearpage

\noindent {\bf Appendix : N=6 Schur function decomposition (continued)}
\vspace{1in}

\begin{table}[h]
\centering
\begin{tabular}{|c|r||c|r||c|r|} \hline
\{$\mu$\}&$a_{\{\mu\}}$  &\{$\mu$\}&$a_{\{\mu\}}$  &\{$\mu$\}&$a_{\{\mu\}}$
\\ \hline
\{9, 7, 7, 4, 2, 1\}&-36&\{9, 6, 6, 5, 3, 1\}&-81&\{8, 8, 8, 2, 2, 2\}&225\\
\{8, 8, 7, 4, 2, 1\}&108&\{8, 7, 6, 5, 3, 1\}&162&\{10, 8, 5, 3, 2, 2\}&27\\
\{10, 7, 5, 5, 2, 1\}&18&\{7, 7, 7, 5, 3, 1\}&162&\{9, 9, 5, 3, 2, 2\}&-81\\
\{9, 8, 5, 5, 2, 1\}&-36&\{8, 6, 6, 6, 3, 1\}&162&\{10, 7, 6, 3, 2, 2\}&-54\\
\{10, 6, 6, 5, 2, 1\}&-54&\{7, 7, 6, 6, 3, 1\}&-486&\{9, 8, 6, 3, 2, 2\}&108\\
\{9, 7, 6, 5, 2, 1\}&72&\{10, 7, 4, 4, 4, 1\}&-36&\{9, 7, 7, 3, 2, 2\}&81\\
\{8, 8, 6, 5, 2, 1\}&54&\{9, 8, 4, 4, 4, 1\}&72&\{8, 8, 7, 3, 2, 2\}&-243\\
\{8, 7, 7, 5, 2, 1\}&-162&\{10, 6, 5, 4, 4, 1\}&72&\{10, 8, 4, 4, 2, 2\}&-6\\
\{9, 6, 6, 6, 2, 1\}&72&\{9, 7, 5, 4, 4, 1\}&-81&\{9, 9, 4, 4, 2, 2\}&18\\
\{8, 7, 6, 6, 2, 1\}&-144&\{8, 8, 5, 4, 4, 1\}&-117&\{10, 7, 5, 4, 2, 2\}&-27\\
\{7, 7, 7, 6, 2, 1\}&360&\{9, 6, 6, 4, 4, 1\}&-117&\{9, 8, 5, 4, 2, 2\}&54\\
\{10, 8, 5, 3, 3, 1\}&-9&\{8, 7, 6, 4, 4, 1\}&234&\{10, 6, 6, 4, 2, 2\}&111\\
\{9, 9, 5, 3, 3, 1\}&27&\{7, 7, 7, 4, 4, 1\}&-81&\{9, 7, 6, 4, 2, 2\}&-162\\
\{10, 7, 6, 3, 3, 1\}&18&\{10, 5, 5, 5, 4, 1\}&-180&\{8, 8, 6, 4, 2, 2\}&-69\\
\{9, 8, 6, 3, 3, 1\}&-36&\{9, 6, 5, 5, 4, 1\}&216&\{8, 7, 7, 4, 2, 2\}&333\\
\{9, 7, 7, 3, 3, 1\}&-27&\{8, 7, 5, 5, 4, 1\}&108&\{10, 6, 5, 5, 2, 2\}&-18\\
\{8, 8, 7, 3, 3, 1\}&81&\{8, 6, 6, 5, 4, 1\}&-324&\{9, 7, 5, 5, 2, 2\}&81\\
\{10, 8, 4, 4, 3, 1\}&27&\{7, 7, 6, 5, 4, 1\}&-288&\{8, 8, 5, 5, 2, 2\}&-153\\
\{9, 9, 4, 4, 3, 1\}&-81&\{7, 6, 6, 6, 4, 1\}&720&\{9, 6, 6, 5, 2, 2\}&-117\\
\{10, 7, 5, 4, 3, 1\}&-36&\{9, 5, 5, 5, 5, 1\}&225&\{8, 7, 6, 5, 2, 2\}&234\\
\{9, 8, 5, 4, 3, 1\}&72&\{8, 6, 5, 5, 5, 1\}&-405&\{7, 7, 7, 5, 2, 2\}&-711\\
\{10, 6, 6, 4, 3, 1\}&-27&\{7, 7, 5, 5, 5, 1\}&90&\{8, 6, 6, 6, 2, 2\}&-36\\
\{9, 7, 6, 4, 3, 1\}&99&\{7, 6, 6, 5, 5, 1\}&675&\{7, 7, 6, 6, 2, 2\}&108\\
\{8, 8, 6, 4, 3, 1\}&-162&\{6, 6, 6, 6, 5, 1\}&-1575&\{10, 8, 4, 3, 3,
2\}&-45\\
\{8, 7, 7, 4, 3, 1\}&-81&\{10, 8, 6, 2, 2, 2\}&-15&\{9, 9, 4, 3, 3, 2\}&135\\
\{10, 6, 5, 5, 3, 1\}&81&\{9, 9, 6, 2, 2, 2\}&45&\{10, 7, 5, 3, 3, 2\}&81\\
\{9, 7, 5, 5, 3, 1\}&-162&\{10, 7, 7, 2, 2, 2\}&45&\{9, 8, 5, 3, 3, 2\}&-162\\
\{8, 8, 5, 5, 3, 1\}&81&\{9, 8, 7, 2, 2, 2\}&-90&\{10, 6, 6, 3, 3, 2\}&-18\\
\hline
\end{tabular}
\end{table}

\newpage
\clearpage

\noindent {\bf Appendix : N=6 Schur function decomposition (continued)}
\vspace{1in}

\begin{table}[h]
\centering
\begin{tabular}{|c|r||c|r||c|r|} \hline
\{$\mu$\}&$a_{\{\mu\}}$  &\{$\mu$\}&$a_{\{\mu\}}$  &\{$\mu$\}&$a_{\{\mu\}}$
\\ \hline
\{9, 7, 6, 3, 3, 2\}&-81&\{7, 6, 6, 5, 4, 2\}&-1215&\{10, 5, 4, 4, 4,
3\}&-420\\
\{8, 8, 6, 3, 3, 2\}&333&\{6, 6, 6, 6, 4, 2\}&405&\{9, 6, 4, 4, 4, 3\}&720\\
\{8, 7, 7, 3, 3, 2\}&-54&\{8, 5, 5, 5, 5, 2\}&450&\{8, 7, 4, 4, 4, 3\}&-180\\
\{10, 7, 4, 4, 3, 2\}&72&\{7, 6, 5, 5, 5, 2\}&-900&\{9, 5, 5, 4, 4, 3\}&675\\
\{9, 8, 4, 4, 3, 2\}&-144&\{6, 6, 6, 5, 5, 2\}&2250&\{8, 6, 5, 4, 4,
3\}&-1215\\
\{10, 6, 5, 4, 3, 2\}&-144&\{10, 8, 3, 3, 3, 3\}&105&\{7, 7, 5, 4, 4, 3\}&270\\
\{9, 7, 5, 4, 3, 2\}&162&\{9, 9, 3, 3, 3, 3\}&-315&\{7, 6, 6, 4, 4, 3\}&345\\
\{8, 8, 5, 4, 3, 2\}&234&\{10, 7, 4, 3, 3, 3\}&-180&\{8, 5, 5, 5, 4, 3\}&-900\\
\{9, 6, 6, 4, 3, 2\}&234&\{9, 8, 4, 3, 3, 3\}&360&\{7, 6, 5, 5, 4, 3\}&1800\\
\{8, 7, 6, 4, 3, 2\}&-468&\{10, 6, 5, 3, 3, 3\}&45&\{6, 6, 6, 5, 4, 3\}&-720\\
\{7, 7, 7, 4, 3, 2\}&162&\{9, 7, 5, 3, 3, 3\}&162&\{7, 5, 5, 5, 5, 3\}&1050\\
\{10, 5, 5, 5, 3, 2\}&45&\{8, 8, 5, 3, 3, 3\}&-711&\{6, 6, 5, 5, 5, 3\}&-3150\\
\{9, 6, 5, 5, 3, 2\}&108&\{9, 6, 6, 3, 3, 3\}&-81&\{10, 4, 4, 4, 4, 4\}&945\\
\{8, 7, 5, 5, 3, 2\}&-351&\{8, 7, 6, 3, 3, 3\}&162&\{9, 5, 4, 4, 4, 4\}&-1575\\
\{8, 6, 6, 5, 3, 2\}&-162&\{7, 7, 7, 3, 3, 3\}&-90&\{8, 6, 4, 4, 4, 4\}&405\\
\{7, 7, 6, 5, 3, 2\}&801&\{10, 6, 4, 4, 3, 3\}&270&\{7, 7, 4, 4, 4, 4\}&45\\
\{7, 6, 6, 6, 3, 2\}&-180&\{9, 7, 4, 4, 3, 3\}&-486&\{8, 5, 5, 4, 4, 4\}&2250\\
\{10, 6, 4, 4, 4, 2\}&-90&\{8, 8, 4, 4, 3, 3\}&108&\{7, 6, 5, 4, 4, 4\}&-720\\
\{9, 7, 4, 4, 4, 2\}&162&\{10, 5, 5, 4, 3, 3\}&-75&\{6, 6, 6, 4, 4, 4\}&225\\
\{8, 8, 4, 4, 4, 2\}&-36&\{9, 6, 5, 4, 3, 3\}&-288&\{7, 5, 5, 5, 4, 4\}&-3150\\
\{10, 5, 5, 4, 4, 2\}&270&\{8, 7, 5, 4, 3, 3\}&801&\{6, 6, 5, 5, 4, 4\}&945\\
\{9, 6, 5, 4, 4, 2\}&-324&\{8, 6, 6, 4, 3, 3\}&-108&\{6, 5, 5, 5, 5, 4\}&4725\\
\{8, 7, 5, 4, 4, 2\}&-162&\{7, 7, 6, 4, 3, 3\}&-201&\{5, 5, 5, 5, 5,
5\}&-10395\\
\{8, 6, 6, 4, 4, 2\}&666&\{9, 5, 5, 5, 3, 3\}&90& & \\
\{7, 7, 6, 4, 4, 2\}&-108&\{8, 6, 5, 5, 3, 3\}&324& & \\
\{9, 5, 5, 5, 4, 2\}&-405&\{7, 7, 5, 5, 3, 3\}&-1422& & \\
\{8, 6, 5, 5, 4, 2\}&567&\{7, 6, 6, 5, 3, 3\}&270& & \\
\{7, 7, 5, 5, 4, 2\}&324&\{6, 6, 6, 6, 3, 3\}&45& & \\ \hline
\end{tabular}
\end{table}


\newpage
\clearpage

\begin{figure}
    \caption{The exact density profiles $\hat{\rho }(r)$, scaled as in
(\protect{\ref{scaling}}) so that the sample edge is at $r=1$ for all $N$, of
the $1\over 3$ filled Laughlin state for N=2,3,4,5,6, together with the
perfectly uniform profile.}
    \label{densityplot}
\end{figure}


\begin{figure}
\centering
    \caption{The density profiles of the maximally distributed states
(\protect{\ref{spread}}) for $N=10,100,1000$ electrons. The plateau is {\it
exactly} at density $1\over 3$.}
    \label{uniplot}
\end{figure}


\begin{figure}
\centering
    \caption{The density profiles of the states (\protect{\ref{<bunched}}),
showing their conribution to the 'boundary hump' for the $N=2,3,4,5,6$ electron
cases.}
    \label{humpplot}
\end{figure}


\begin{figure}
\centering
    \caption{The density profiles of the dominant states
(\protect{\ref{<spreadmu}}) for $N=10,100,1000$ electrons. As in Figure
\protect{\ref{uniplot}}, the plateau is exactly at density $1\over 3$.}
    \label{domplot}
\end{figure}


\begin{figure}
\centering
    \caption{The pair correlation function for the $1\over 3$ filled $N=6$
Laughlin state, as computed in equation (\protect{\ref{gcorr}}), together with
the approximate correlation function (\protect{\ref{gapprox}}) obtained by
retaining a dominant Slater state in (\protect{\ref{gcorr}}).}
    \label{corrplot}
\end{figure}



\end{document}